\documentclass[aps,prl,reprint,twocolumn,english,superscriptaddress,longbibliography]{revtex4-2}
\usepackage[utf8]{inputenc}
\usepackage{graphicx}
\usepackage{bm}
\usepackage{booktabs} 
\usepackage{graphicx}
\usepackage{siunitx}
\usepackage{physics}
\usepackage[caption=false]{subfig}
\usepackage{floatrow}
\usepackage{xargs}  
\usepackage[pdftex,dvipsnames]{xcolor}
\usepackage{eqparbox}

\begin{document}

\title{Realization of a multi-output quantum pulse gate for decoding high-dimensional temporal modes of single-photon states}
\author{Laura Serino}
\email[]{laura.serino@upb.de}
\author{Jano Gil-Lopez} 
\author{Michael Stefszky} 
\author{Raimund Ricken} 
\author{Christof Eigner} 
\author{Benjamin Brecht}
\author{Christine Silberhorn}
\affiliation{Paderborn University, Integrated Quantum Optics, Institute for Photonic Quantum Systems (PhoQS), Warburgerstr.\ 100, 33098 Paderborn, Germany}

\begin{abstract}
Temporal modes (TMs) of photons provide an appealing high-dimensional encoding basis for quantum information. While techniques to generate TM states have been established, high-dimensional decoding of single-photon TMs remains an open challenge. In this work, we experimentally demonstrate demultiplexing of five-dimensional TMs of single photons with an average fidelity of $0.96 \pm 0.01$, characterized via measurement tomography. This is achieved using a newly developed device, the multi-output quantum pulse gate (mQPG). We demonstrate a proof-of-principle complete decoder based on the mQPG that operates on any basis from a set of 6 five-dimensional MUBs and is therefore suitable as a receiver for high-dimensional quantum key distribution. Furthermore, we confirm the high-quality operation of the mQPG by performing resource-efficient state tomography with an average fidelity of $0.98 \pm 0.02$.
\end{abstract}


\maketitle

\section{Introduction}
As increasingly more applications for quantum technologies continue to be found, the need to develop a device that enables highly efficient quantum communication (QC) becomes critical \cite{kimble08, obrien09, wang20}. Photons are ideally suited for this task due to transmission at the speed of light, intrinsically low decoherence and to their high-dimensional spatial and time-frequency degrees of freedom. These degrees of freedom provide high-dimensional alphabets that allow one to encode more information per photon, leading to important advantages for QC applications, including the higher level of security and efficiency provided by high-dimensional quantum key distribution (HD-QKD) with respect to its binary counterpart \cite{sheridan10}.

Arguably, the most explored high-dimensional photonic degree of freedom is the spatial one, with particular focus on the orbital angular momentum (OAM) of light. One of the main advantages of this encoding alphabet is the possibility to generate and detect states using time-invariant operations \cite{berkhout10, huang18, mirhosseini15}. However, this advantage comes with an important drawback, as OAM states are inherently incompatible with existing fiber optics networks and easily degrade in free space transmission \cite{cozzolino19}.

The time-frequency degree of freedom of photons overcomes this limitation. The standard alphabet based on this degree of freedom is given by time and frequency bins, which can be conveniently generated from an integrated source \cite{reimer16, kues19}. The manipulation of time-frequency bins and their superpositions is possible with interferometric systems \cite{humphreys13} or with a combination of phase modulators and pulse shapers \cite{lu18a, lu18b}. Scaling either of those approaches comes with important challenges, and experimental efforts have been limited to low-dimensional systems \cite{islam17, vagniluca20}.

To avoid this impediment, we can exploit the time-frequency degree of freedom of light through temporal modes (TMs), i.e. field-orthogonal wave-packet modes. Since TMs span an infinite-dimensional Hilbert space, they can represent any arbitrary time-frequency state of single photons;
of particular importance is that they form a natural basis to describe photons generated through ultrafast parametric down-conversion.
TMs are characterized by robustness against fiber dispersion and a higher packing density with respect to frequency bins, as they can exploit the full frequency space without the need for guard bands \cite{brecht15, raymer20}.
These properties render TMs a valuable resource not only for QC protocols, but also for several other applications, from quantum enhanced spectroscopy and metrology \cite{schlawin16, ansari21, mazelanik22} to quantum memories \cite{nunn07, afzelius09, zheng15} and deterministic photonic quantum gates \cite{harder13, ralph15, gao19}.

In order to fully reap the benefits of high-dimensional encoding, a TM-based QC scheme requires the generation, manipulation and simultaneous detection of multiple TMs of single photons.
Single-photon operation is a requirement as quantum light is employed in a wide variety of interfacing and communication protocols.
Generation of single photons with a well-defined TM state has been successfully achieved through ultrafast parametric down-conversion \cite{ansari18a, shi08, mosley08}.
Manipulation and detection of single-photon TMs, on the other hand, have been limited to a single TM at a time. These demonstrations were obtained employing a quantum pulse gate (QPG) \cite{brecht14, ansari18b, reddy18}, a TM-selective device based on integrated sum-frequency generation that by design is limited to single-output operation.

In this work, we demonstrate high-dimensional single-photon TM decoding using a multi-output QPG (mQPG). The mQPG is a newly developed device that employs a custom poling structure to project a single-photon-level input signal onto all the elements of a chosen high-dimensional TM alphabet (or their superpositions) and map the results of the projections onto different output frequencies. We then make use of a single-photon spectrograph to read out the output frequency for each input photon, hence providing the projection result as a ``click'' in the corresponding output channel.
We demonstrate that our device is compatible with single-photon-level input states from a five-dimensional Hilbert space, and characterize its performance through a quantum measurement tomography. Thus we showcase a proof-of-principle complete HD-QKD decoder based on the mQPG that can work with any basis from a set of 6 five-dimensional mutually unbiased bases (MUBs). To further confirm the capabilities of our device, we use the reconstructed positive-operator-valued measure (POVM) from the measurement tomography to facilitate a resource-efficient state tomography with an average fidelity of $0.98 \pm 0.02$.
Furthermore, we describe the necessary improvements to scale the decoder scheme to perform high-quality measurements in even higher dimensions.

We note that, although preliminary work towards expanding frequency conversion to multi-output operation has been demonstrated \cite{silver19, chou99}, these implementations were inherently incompatible with single-photon-level input states. Our device allows one to fully exploit the higher information capacity provided by high-dimensional encoding in single photons, granting faster transmission of information and enabling high-dimensional quantum key distribution (HD-QKD) protocols \cite{sheridan10}. Moreover, the mQPG substantially accelerates measurement times for applications that require projections onto a large number of TMs, such as quantum state tomography \cite{ansari18b}.

\section{Device and process engineering}
\label{sec:device}
\begin{figure}
    \includegraphics[]{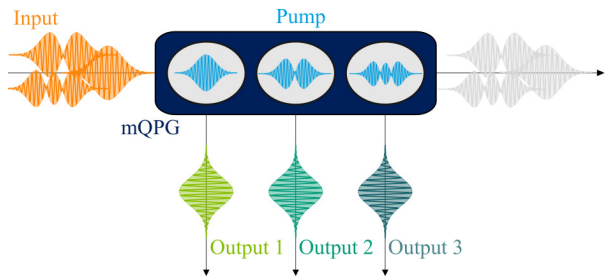}
    \caption{\label{fig:concept_mqpg} Working principle of the mQPG: the input photons are upconverted to different frequencies based on their temporal mode.}
\end{figure}

A high-dimensional decoder must allow the user to perform a simultaneous high-quality projection of the input state onto all the elements of a user-chosen basis (Figure \ref{fig:concept_mqpg}). 
The mQPG achieves this goal through type-II sum-frequency generation (SFG) in periodically poled titanium-indiffused lithium niobate waveguides. 
Its operation in frequency space is described by a transfer function $G$, which maps the input frequencies $\nu_\mathrm{in}$ onto the sum frequencies $\nu_\mathrm{out}$.

To better understand the nature of the mQPG process, we can first describe the simpler single-output process of the quantum pulse gate (QPG) \cite{brecht14}. The QPG is a device that selectively upconverts a specific TM component from an input state with an efficiency of up to 87.7\%. Such a high conversion efficiency is in principle also achievable by the mQPG process.

The transfer function $G_0$ of the QPG is the product of the pump function $\alpha_0$, that describes energy conservation, and the phase-matching function $\Phi_0$:
\begin{equation}
    G_0(\nu_\mathrm{in}, \nu_\mathrm{out}) = \alpha_0(\nu_\mathrm{in}, \nu_\mathrm{out}) \cdot \Phi_0(\nu_\mathrm{in}, \nu_\mathrm{out}) \,.
\end{equation}
The SFG process of the QPG (and mQPG alike) is dispersion-engineered for group-velocity matching of the pump and input fields, which causes the phase-matching function to be independent of the input frequency ($\Phi_0 \simeq \Phi_0(\nu_\mathrm{out})$).
Under the condition that the input and pump bandwidths are significantly broader than the phase matching bandwidth, the transfer function $G_0$ becomes separable into a pair of input and output functions \cite{donohue18}, meaning that the QPG can perform selective upconversion of a specific TM.
As a consequence, the intensity of the SFG light is proportional to the overlap between the complex spectral amplitudes of the input and pump pulses. The latter can be tailored through spectral shaping, allowing one to select the desired TM, transmitting the orthogonal TM
components unperturbed.

\begin{figure}
    \captionsetup[subfloat]{position=top}
    \subfloat[\label{fig:tf_sim} Simulation]{%
        \raggedleft
        \includegraphics[]{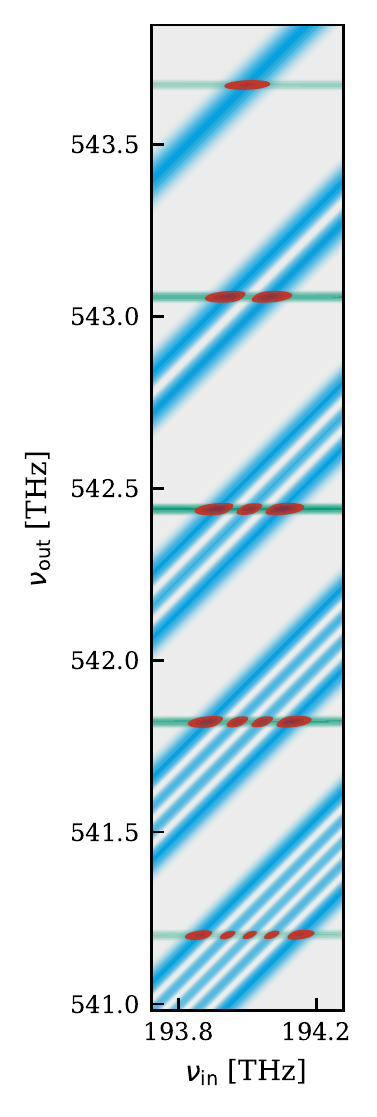}
    }
    \hfill
    \subfloat[\label{fig:tf_exp} Experiment]{%
        \raggedleft
        \includegraphics[]{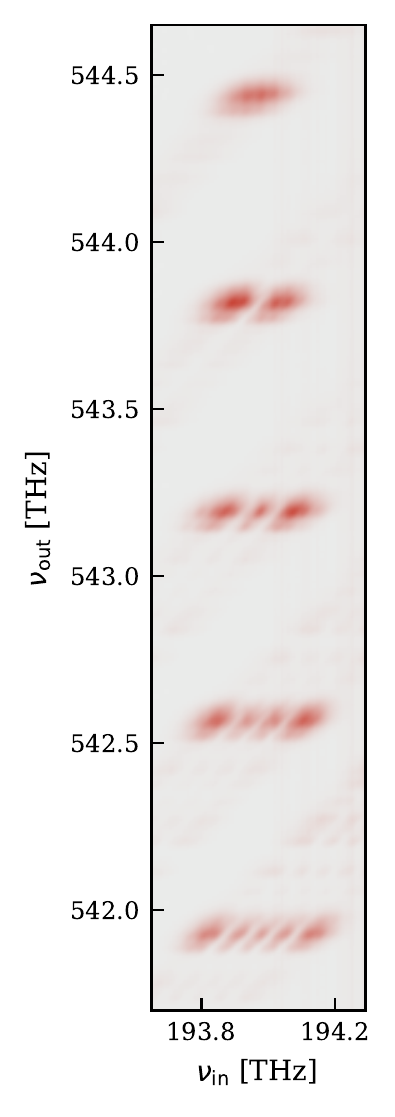}
    }
    \caption{\label{fig:tf}
             Figure \ref{fig:tf_sim} shows the simulated transfer function $G(\nu_\mathrm{in}, \nu_\mathrm{out})$ (red) given by the product of a five-peak phase-matching function $\Phi(\nu_\mathrm{in}, \nu_\mathrm{out}) \simeq \Phi(\nu_\mathrm{out})$ (green) and a five-peak shaped pump spectrum $\alpha(\nu_\mathrm{in}, \nu_\mathrm{out})$ (blue). The pump peaks are shaped as the first five Hermite-Gaussian modes and only the intensity of the named quantities is represented. Figure \ref{fig:tf_exp} shows the experimental transfer function, which resembles well the simulation. Upon closer inspection, the transfer function displays a small amount of distortion below the main peaks, which can be attributed to fabrication defects \cite{santandrea19}. These imperfections can be removed applying spectral filtering to each output channel.}
\end{figure}

The inherent limitation of the QPG to a single output channel limits the detection to one single TM at a time, making this device unsuitable for HD-QKD, which requires the detection of any element of the chosen basis at every single shot of the communication. In contrast, the multi-channel nature of the mQPG renders it suitable for high-dimensional demultiplexing of many TMs into distinct output channels. Each channel $j$ of the mQPG acts as a distinct QPG with its own transfer function $G_j$ and maps a specific TM mode into the corresponding output frequency. The multi-peak phase-matching function $\Phi = \sum_j{\Phi_j}$ of the mQPG is generated through the alternation of periodically poled and unpoled regions along the waveguide (Figure \ref{fig:pm}). The design framework of the mQPG, illustrated in detail in the Supplementary Material, allows one to freely select the spectral parameters such as the inter-peak distance and effective number of peaks.
One can then shape a pump spectrum $\alpha = \sum_j{\alpha_j}$ with as many peaks as $\Phi$ in order to create a multi-output transfer function such as the one illustrated in Figure \ref{fig:tf}.


For this demonstration, we use a five-peak mQPG waveguide and generate a five-output transfer function to facilitate operation in a five-dimensional space. The experimental transfer function, shown in Figure \ref{fig:tf_exp}, matches its simulated counterpart (red region in Figure \ref{fig:tf_sim}). The technical specifications of the mQPG waveguide used in this experiment are explained in detail in the Supplementary Material.

\begin{figure}
    \captionsetup[subfloat]{position=top}
    \centering
    \subfloat[\label{fig:superpoling}]{%
        \includegraphics[]{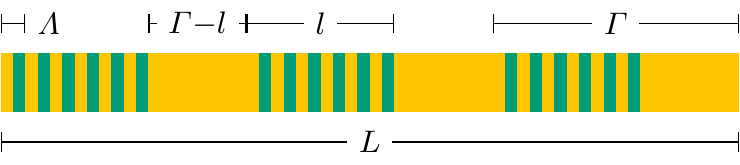}
    }
    \vfill
    \subfloat[\label{fig:pm_1D}]{%
        \includegraphics[]{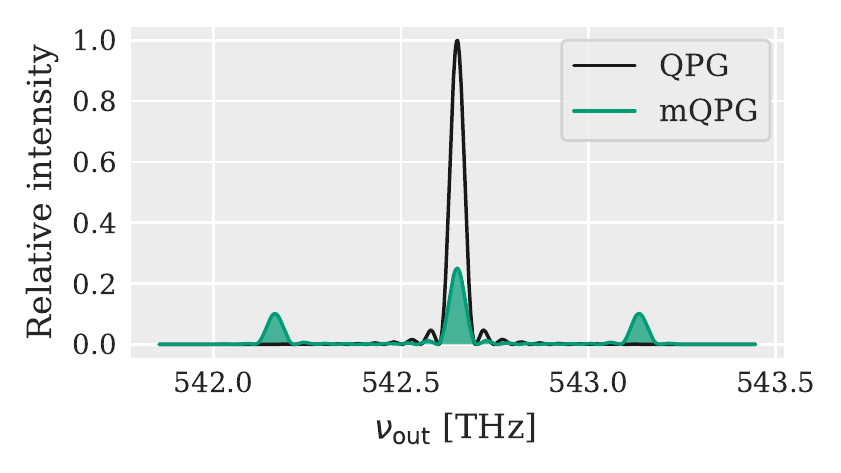}
    }
    \caption{\label{fig:pm} Figure \ref{fig:superpoling} shows the schematic of the super-poling pattern of an mQPG waveguide: regions of length $l$ poled with period $\varLambda$ are alternating with unpoled regions of length $\varGamma-l$, where $\varGamma$ is the period of this alternation. The regions corresponding to the standard nonlinearity coefficient of the waveguide are shown in yellow, whereas those corresponding to the reversed nonlinearity coefficient are represented in green. Figure \ref{fig:pm_1D} shows a simulation of the phase-matching intensity spectrum of an mQPG (green) compared to that of a QPG (black). We note that the lower efficiency of an mQPG can be compensated by using higher pump powers.}
\end{figure}

\section{Tomography theory}
In order to quantify the quality of the high-dimensional decoding of the mQPG, we perform a quantum measurement tomography \cite{lundeen09}. For this purpose, we first introduce the mathematical description of the mQPG operation.
We can describe the multi-output decoding as a positive operator-valued measure (POVM) $\{\pi^\gamma\}$, which corresponds to the complete measurement basis comprising all the channels of the mQPG. Each individual channel of the mQPG projects the input state onto a user-chosen TM, $\gamma$. Consequently, each channel corresponds to a POVM element, i.e.\ measurement operator, $\pi^\gamma = \sum_{ij}{m^\gamma_{ij}\ket{i}\bra{j}}$ (ideally $\ket{\gamma}\bra{\gamma}$), with $i$ and $j$ elements of the fundamental TM basis.
We choose to work in a $d$-dimensional Hilbert space, where $d$ matches the number of channels of the decoder. We select the decoder TMs $\{\gamma\}$ to form a basis for the aforementioned space.

For each mQPG channel, the probability of SFG conversion of a pure input state $\rho^\xi = \ket{\xi}\bra{\xi}$ is \cite{ansari17}
\begin{equation}
    p^{\gamma\xi} = \Tr{\rho^\xi\pi^\gamma} \,.
\end{equation}

The aim of measurement tomography is to probe the decoder with a full set of input states $\rho^\xi$ in order to reconstruct the full POVM $\{\pi^\gamma\}$.
In this work, we employ a set of probe states containing all the elements of the $d+1$ mutually unbiased bases (MUBs) of our $d$-dimensional Hilbert space, demonstrating the compatibility of the mQPG with quantum communication protocols based on MUBs, including HD-QKD.

\section{Experiment}
\label{sec:experiment}
\begin{figure}
    \centering
    \includegraphics[width=\linewidth]{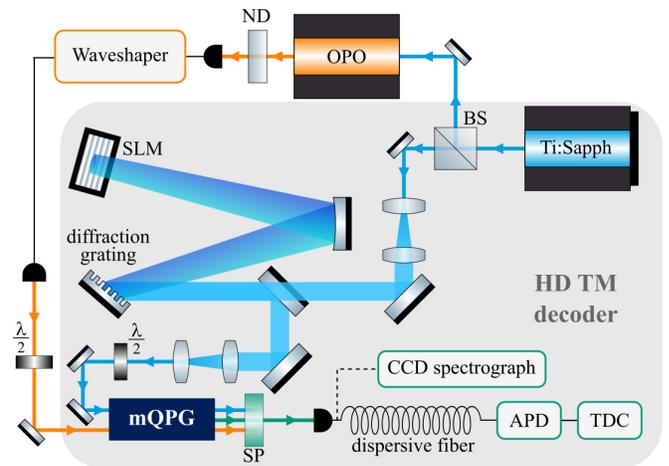}
    \caption{Schematic of the experimental setup. The blue line indicates the path of the pump beam, the orange line corresponds to the input signal and the green line shows the output field of the mQPG. The gray area indicates the components of our high-dimensional temporal-mode decoder (for more information, see the text).}
    \label{fig:setup}
\end{figure}

The schematic of the experimental setup is shown in Figure \ref{fig:setup}. We start from ultrashort Ti:Sapphire laser pulses with a spectrum centered around $\lambda_\mathrm{p, 0} = \SI{860}{\nano\meter}$ (\SI{349}{\tera\hertz}) and a repetition rate of \SI{80}{\mega\hertz}. A portion of the beam is directed to a home-built 4-f line based on a spatial light modulator (SLM) \cite{monmayrant10} with a resolution $\delta\nu_\mathrm{shaper, p} = \SI{10}{\giga\hertz}$ to shape the amplitude and phase of its complex spectrum
in order to prepare the pump states. The pulse is carved into five peaks with centers separated by $\Delta\nu_\mathrm{sep} = \SI{0.63}{\tera\hertz}$ , each shaped as an element of a five-dimensional TM basis. For this experiment, we choose Hermite-Gaussian (HG) modes and their superpositions, as they provide a good approximation of the natural modes of parametric down-conversion processes \cite{law00}. 
The remaining part of the pulse train pumps an optical parametric oscillator that generates pulses centered at $\lambda_\mathrm{in, 0} = \SI{1545}{\nano\meter}$ (\SI{194}{\tera\hertz}). The beam is attenuated to the single photon level with a mean photon number per pulse lower than 0.1 ($0.097\pm0.001$ photons per pulse) and shaped using a commercial waveshaper with a resolution of $\delta\nu_\mathrm{shaper, in} = \SI{1}{\giga\hertz}$ 
to prepare the input state. 
The shaping parameters for the input beam and each pump peak are chosen such that the FWHM of the fundamental HG mode is $\Delta\nu_\mathrm{p} = \Delta\nu_\mathrm{in} = \SI{0.14}{\tera\hertz}$.
Both beams are then coupled into the mQPG waveguide, each with a coupling efficiency of approximately 60\%. The waveguide is designed to be spatially single mode for telecom light, and particular care is taken to ensure the pump field is coupled in the fundamental spatial mode.

The multi-output SFG process of the mQPG generates output fields at multiple frequencies (each defining an output channel) around \SI{552}{\nano\meter} (\SI{543}{\tera\hertz}) based on the TM state of the input field. Effectively, the mQPG projects the input TM onto the chosen TM basis and maps the results to the corresponding output frequencies. The internal conversion efficiency is approximately 5\%, here limited by the pump pulse energy available in this experiment.
The output fields are separated from the residual pump and input fields using a dichroic mirror and then fiber-coupled and measured with a commercial CCD spectrograph (Andor Shamrock 500i) with a resolution of \SI{30}{\giga\hertz}.

\begin{figure*}
    \centering
    \includegraphics[]{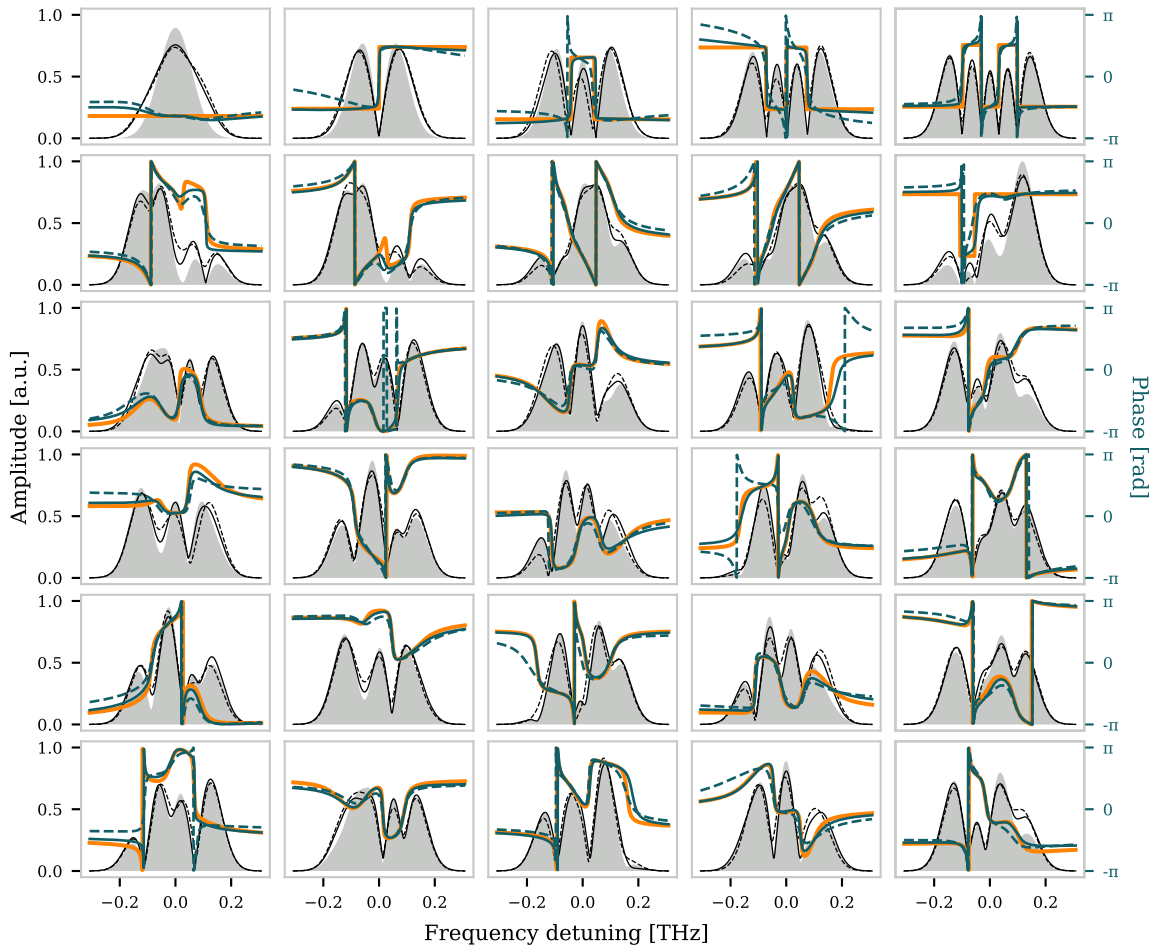}
    \caption{First eigenmodes of the POVM elements (5 for each of the 6 MUBs). Shaded areas and orange lines show the ideal amplitude and phase; black and blue lines show the same values for experimental data, with solid and dashed line corresponding to CCD and ToF spectrograph respectively.}
    \label{fig:det_tomo}
\end{figure*}

To perform the quantum measurement tomography, we probe the decoder with input states from a set comprising all the 30 elements of the 6 MUBs of our Hilbert space.
For each measurement we run 3 acquisitions, each with an integration time of \SI{10}{\second} and an average count rate of \SI{470}{\hertz}. 
We then calculate the experimental output probabilities for each channel $p^{\gamma\xi}$ and we use them to reconstruct the POVM elements $\pi^\gamma$ through a weighted least-squares fit:
\begin{equation}
    \min_{\pi^\gamma}{\sum_\xi{\frac{\left|p^{\gamma\xi} - \Tr{\rho^\xi\pi^\gamma}\right|^2}{p^{\gamma\xi}}}} \,,
\end{equation}
where we constrain $\pi^\gamma$ to be Hermitian and positive semidefinite. 
The first eigenmodes of the reconstructed POVM elements are presented in Figure \ref{fig:det_tomo}. One can see that they very closely match the ideal POVMs both in amplitude and phase.

We quantify the quality of the decoder by calculating the purities of the POVM elements
\begin{equation}
    \mathcal{P^\gamma} = \frac{\Tr{(\pi^\gamma)^2}}{\Tr{\pi^\gamma}^2}
\end{equation}
and the fidelities when compared to the ideal operators
\begin{equation}
    \mathcal{F^\gamma} = \sqrt{ \frac{\expval{\pi^\gamma}{\gamma}}{\Tr{\pi^\gamma} }} \,.
\end{equation}
In an ideal system, both these values are equal to 1. The average experimental results with their respective standard deviations are listed in Table \ref{tab:det_tomo}, left column. The high average fidelity and purity indicate the remarkably good quality of the mQPG measurements.

\begin{table}
    \setlength{\tabcolsep}{6pt}
    \centering
    \begin{tabular}{lcc}
        \toprule
         & CCD spectrograph & ToF spectrograph \\
        \midrule
        Fidelity & $0.956 \pm 0.014$ & $0.810 \pm 0.046$\\
        Purity & $0.885 \pm 0.036$ & $0.552 \pm 0.087$\\
        \bottomrule
    \end{tabular}
    \caption{Average fidelities and purities of the reconstructed POVM elements in $d=5$ with the corresponding standard deviation.}
    \label{tab:det_tomo}
\end{table}

\section{HD-QKD decoder demonstration}
To demonstrate a proof-of-principle HD-QKD decoder, we need to perform photon counting. For this reason, we replace the CCD spectrograph with a home-built time-of-flight (ToF) fiber-assisted single-photon spectrograph \cite{avenhaus09, marsili13}.
This system exploits the chromatic group velocity dispersion (GVD) of a single-mode fiber to apply varying delays to the different frequency components of the pulses, effectively mapping each frequency component to a different arrival time. 
The arrival times are then measured using an avalanche photodiode (APD) combined with a time-to-digital converter. A calibration of the system then allows one to calculate the frequency corresponding to each arrival time, recovering the information on the detected TM.
The simple structure of the ToF spectrograph makes it cost effective and versatile.
The ToF setup used in this experiment has a resolution of \SI{0.3}{\tera\hertz} and introduces \SI{20}{\decibel} of losses. These values are limited by our availability of low-loss high-GVD fibers for visible light. The high losses do not pose a fundamental problem for the operation of our decoder, however, they limit the maximum distance at which communication is still possible.

As a first characterization of the complete HD-QKD decoder, we perform a new measurement tomography. For each measurement we run 30 acquisitions, each with an integration time of \SI{10}{\second} and an average count rate of \SI{40}{\hertz}.
The characterization results, reported in Table \ref{tab:det_tomo}, show a good average fidelity. The good fidelity is apparent in the agreement between the first eigenmodes of the reconstructed POVMs and the ideal ones (Figure \ref{fig:det_tomo}), which show a similarly good quality for all the 6 MUBs.
The low average purity of the reconstructed POVMs, however, is an indicator of multi-modedness in the decoder, which results in output clicks corresponding to TMs that are not present in the input state. The discrepancy with the high intrinsic purity of the mQPG indicates that the loss of quality is to be traced back uniquely to the difference between the ToF spectrograph and the CCD spectrograph. As a matter of fact, the limited resolution of the ToF spectrograph is not sufficient to completely filter out the phase-matching imperfections of the mQPG (visible in Figure \ref{fig:tf_exp}), which then introduce a multi-mode behavior in the system.

\begin{figure}
    \centering
    \includegraphics{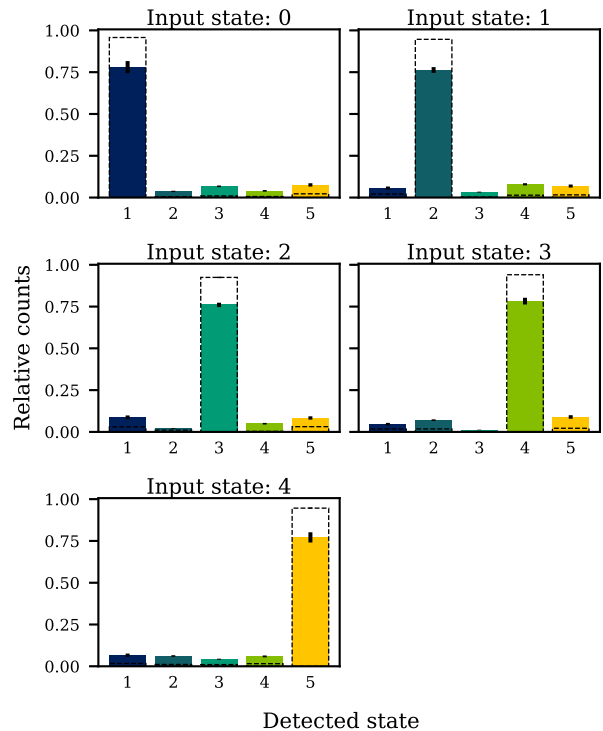}
    \caption{Relative counts obtained when projecting the five elements of one MUB onto the same MUB with the complete decoder (filled bars) and internal performance of the mQPG (dashed lines).}
    \label{fig:counts}
\end{figure}

Despite the loss of quality due to the spectrograph, the decoder shows good TM-demultiplexing behavior, which makes it suitable as a proof-of-principle receiver for HD-QKD. To demonstrate this, we extract the count rates relative to the projection of each input state onto the corresponding MUB from the same set of experimental data used for the tomography. We then calculate the average selectivity per MUB as the ratio of the number of correct counts to the total number of counts.
Our results show an average selectivity per basis ranging from 61\% to 78\%, which demonstrate a clear mode-selective behavior, but indicate that there is room for improvement. This value is in fact lower than the internal average selectivity of the mQPG $\mathcal{S}=92\%$ (measured with the CCD spectrograph). The discrepancy is once again explained by the low resolution of the ToF spectrograph employed in this experiment, which hinders proper discrimination of the counts. This becomes evident when comparing the relative counts measured by the decoder in one of the 6 MUBs to the intrinsic performance of the mQPG (Fig. \ref{fig:counts}). This observation suggests that, although the current implementation of the mQPG-based decoder is already up to the task of TM demultiplexing, a high-resolution single-photon-resolving spectrograph will allow one to take full advantage of the high measurement quality provided by the mQPG.

\section{Beyond QKD: state tomography}
In this section, we demonstrate that the measurement tomography of the mQPG allows us to overcome its shortcomings for applications that require measurements that integrate over time, such as state tomography. For this purpose, we perform resource-efficient single-photon state tomography with the mQPG and we show the improvement provided by the reconstruction of the experimental POVMs.
We prepare 25 random pure input states $\rho$ from our five-dimensional Hilbert space, and measure them using the 6 MUBs as decoder bases. We perform the frequency-resolved detection with the CCD camera, running 3 acquisitions for each measurement, each with an integration time of \SI{10}{\second} and an average count rate of \SI{470}{\hertz}. We reconstruct the input states through a weighted least-squares fit:
\begin{equation}
    \min_{\rho}{\sum_\xi{\frac{\left|p^{\gamma} - \Tr{\rho\pi^\gamma}\right|^2}{p^{\gamma}}}} \,,
\end{equation}
where $p^{\gamma}$ are the output probabilities of the different channels and $\rho$ is constrained to be Hermitian and positive semidefinite. We stress that, owing to the high dimensionality of our decoder, it is sufficient to set the pump states to the desired basis to project the input state onto all its elements at the same time within a single measurement. This means that we perform only 6 measurements to obtain counts on all 30 possible pump states. We highlight that the high number of required measurements constituted the main limitation to dimension scalability for the QPG \cite{ansari17} and that the mQPG overcomes this limitation.

\begin{figure}
    \centering
    \includegraphics[]{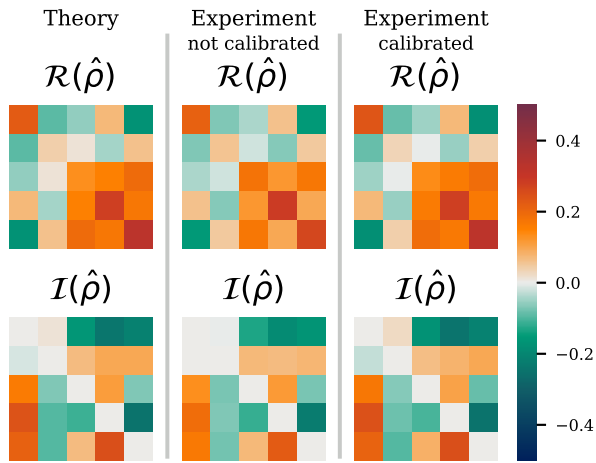}
    \caption{Example of tomography of a random input state (measured with the CCD spectrograph). The plots correspond to the real (top) and imaginary (bottom) parts of the density matrix of, from left to right: the original input state, the raw reconstructed state, and the state reconstructed with the calibration information obtained from the measurement tomography.}
    \label{fig:state_tomo}
\end{figure}

We reconstruct the input state first assuming ideal POVMs $\{\ket{\gamma}\bra{\gamma}\}$ (``raw'' state tomography), and then considering the measured POVMs $\{\pi^\gamma\}$ (``calibrated'' state tomography). The results are summarized in Table \ref{tab:state_tomo}, and an example of input state reconstruction is provided in Figure \ref{fig:state_tomo}. The calibrated state tomography significantly improves fidelity and purity of the reconstructed states. 
This can be explained considering that the main source of error in our system is the residual cross-talk deriving from the phase-matching imperfections, which cannot be completely eliminated even with the high resolution of the CCD spectrograph. This cross-talk is well characterized through the measurement tomography, therefore in the calibrated state tomography we exploit this information in order to correct for the induced errors.


\begin{table}
    \setlength{\tabcolsep}{6pt}
    \centering
    \begin{tabular}{lcc}
        \toprule
        & Raw & Calibrated\\
        \midrule
        Fidelity & $0.941 \pm 0.019$ & $0.983 \pm 0.016$\\
        Purity & $0.816 \pm 0.062$ & $0.941 \pm 0.056$\\
        \bottomrule
    \end{tabular}
    \caption{Average fidelities and purities of the reconstructed input states in $d=5$ with the corresponding standard deviation.}
    \label{tab:state_tomo}
\end{table}

\section{Discussion}
The current implementation of the proof-of-principle decoder for HD-QKD is limited by available components, however, the issues are of technical nature and can be overcome rather easily.
The main limitation is set by the spectrograph system, which is required to be compatible with photon counting at visible wavelengths. Such a device is not yet commercially available, and the in-house-built ToF spectrograph used in this work introduced \SI{20}{\decibel} of losses and had a limited resolution of \SI{0.3}{\tera\hertz}, which decreased the fidelity from the intrinsic value of the mQPG $\mathcal{F} = 0.96 \pm 0.01$ to the resulting decoder fidelity $\mathcal{F'} = 0.81 \pm 0.05$.
The loss of fidelity can be prevented using state-of-the-art components for the spectrograph system. Indeed, a ToF spectrograph based on a state-of-the-art chirped fiber Bragg grating (CFBG) \cite{davis17} and a low-jitter superconducting nanowire single-photon detector (SNSPD) \cite{korzh20} would achieve a resolution of \SI{6}{GHz} for much lower losses, allowing one to fully exploit the high intrinsic fidelity of the mQPG and at the same time increase the efficiency of the decoder.

Moreover, the decoding efficiency can be further improved by increasing the internal conversion efficiency of the mQPG waveguide. This can be done through an optimized tailoring of the phase-matching function \cite{chou99, branczyk11, zhu21} and by increasing the pump pulse energy available in the experiment. To achieve a higher pump energy, one can of course resort to a higher-power laser. However, an appealing solution, with the advantage of on-chip integration, would be to generate each pump peak with a separate chip-based mode-locked laser. Indeed, preliminary devices demonstrated a spectral bandwidth of approximately \SI{0.35}{\tera\hertz} \cite{gordon15}, which matches the ideal bandwidth of each pump peak of the mQPG.

Although the implemented five-channel mQPG carries undeniable benefits for QC applications based on single-photon TMs (which have been until now limited to a mere two-dimensional space) one may wish to push the device to even higher dimensions.
The methods detailed in the Supplementary Material provide the scheme for generating an mQPG waveguide with the desired number of output channels via straightforward modulation of the poling pattern.
The cost of moving to higher dimensions is that each pump peak becomes narrower for a fixed spectral width set by the Ti:Sapphire pump pulse. To prevent a loss in measurement quality, one must either increase the available spectral bandwidth or address the difficulties that arise when operating with narrower and more densely spaced peaks. 
This requires an increased spectrograph resolution to fully discriminate the output channels, an improved shaping resolution, and a narrower phase-matching function (e.g.\ through a longer waveguide \cite{gil-lopez21}) to satisfy the single-modedness condition for each channel \cite{donohue18}.
If these requirements are met, there is no fundamental limitation to the highest dimensionality achievable with this system.

\section{Conclusion}
We demonstrated five-dimensional temporal-mode demultiplexing of single photons using a newly developed device, the mQPG. We characterized its performance through a measurement tomography, obtaining an average fidelity of $0.96 \pm 0.01$ to the ideal POVMs. We then demonstrated a complete HD-QKD decoder for single photons based on the mQPG, which revealed an average fidelity of $0.81 \pm 0.05$, solely limited by the currently available spectrograph technology.
Finally, we exploited the information obtained from the measurement tomography of the mQPG to demonstrate resource-efficient high-quality state tomography with an average fidelity of $0.98 \pm 0.02$. 

These results show that the demonstrated architecture provides a scalable framework for high-dimensional decoding. The purity and fidelity of the device can be easily improved by limiting the phase-matching imperfections and increasing the resolution of the spectrograph. Doing so will result in a decoder with high-performance operation in more than 10 dimensions, leading to even higher information capacity and security in HD-QKD.

Finally, we highlight the versatility of the mQPG, which constitutes one of its main advantages. Due to the fully programmable shaping systems, one can easily work with alternative high-dimensional TM encodings without modifying the experimental setup. This gives one the freedom to explore a wide range of parameters in order to find the optimal solution for different applications. Finally, we emphasize that, independently of the chosen encoding alphabet and of the dimensionality $d$ of the system, the mQPG is always able to work with the full set of $d+1$ MUBs. This makes the mQPG a valuable resource for many QC applications and, furthermore, opens up new opportunities for all TM-based technologies, from quantum spectroscopy and metrology, to quantum memories and deterministic photonic quantum gates.

\section{Acknowledgements}
The authors would like to thank M. Santandrea and V. Ansari for helpful discussions. This research was supported by the EU H2020 QuantERA ERA-NET Cofund in Quantum Technologies project QuICHE.

\bibliography{bibliography}

\end{document}


\title{\Large Supplementary Material\\
\large Realization of a multi-output quantum pulse gate for decoding high-dimensional temporal modes of single-photon states}
\author{Laura Serino}
\email[]{laura.serino@upb.de}
\author{Jano Gil-Lopez} 
\author{Michael Stefszky} 
\author{Raimund Ricken} 
\author{Christof Eigner} 
\author{Benjamin Brecht}
\author{Christine Silberhorn}
\affiliation{Paderborn University, Integrated Quantum Optics, Institute for Photonic Quantum Systems (PhoQS), Warburgerstr.\ 100, 33098 Paderborn, Germany}


\maketitle

\section{Multi-output quantum pulse gate}
In this section, we describe the techniques used to obtain a multi-output quantum pulse gate (mQPG) process for an input signal comprising multiple temporal modes centered at telecommunications wavelengths around \SI{1545}{\nano\meter} (\SI{194}{\tera\hertz}). We start by finding the optimal parameters for a QPG process, i.e., type-II sum-frequency generation (SFG) between two group-velocity matched fields labeled ``pump'' and ``signal''. To achieve group-velocity matching in a non-degenerate process, we exploit the polarization dependency of the group velocity of light in a birefringent crystal. This allows us to find a TM-polarized pump wavelength with the same group velocity as the TE-polarized input field. Next, we achieve the quasi-phase-matching condition for the chosen fields by applying a periodic poling with period $\varLambda$ to the waveguide. With this method, we can identify a QPG process tailored for the chosen input wavelength, in our case a telecom field.

\begin{figure}
    \centering
    \includegraphics[]{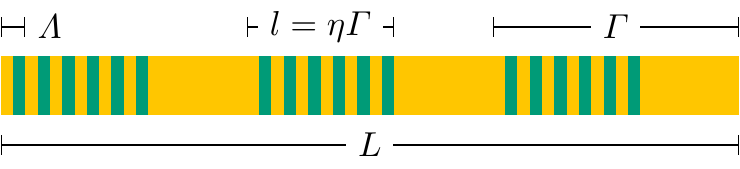}
    \caption{Scheme of the super-poling pattern of an mQPG waveguide. The regions corresponding to the standard nonlinearity coefficient of the waveguide are shown in yellow, whereas those corresponding to the reversed nonlinearity coefficient are represented in green.}
    \label{fig:superpoling} 
\end{figure}

In order to obtain the mQPG, we need to expand the QPG process to multiple output channels. We note that, due to the properties of the QPG, each output channel would correspond to one phase-matching peak. Previous work \cite{chou99, silver19} has demonstrated that multiple phase-matching peaks are attainable through a periodic modulation of the poling pattern. We therefore modify the poling structure of a QPG waveguide by applying a modulation with period $\varGamma$ (``super-poling period'') and an asymmetric duty cycle $\eta\in [0, 1]$, obtaining poled regions of length $l=\eta\varGamma$ alternating with unpoled regions of length $(1-\eta)\varGamma$ (Figure \ref{fig:superpoling}). We can calculate the new phase-matching function analytically to show how this modulation generates multiple output peaks.

\begin{figure}
    \centering
    \includegraphics{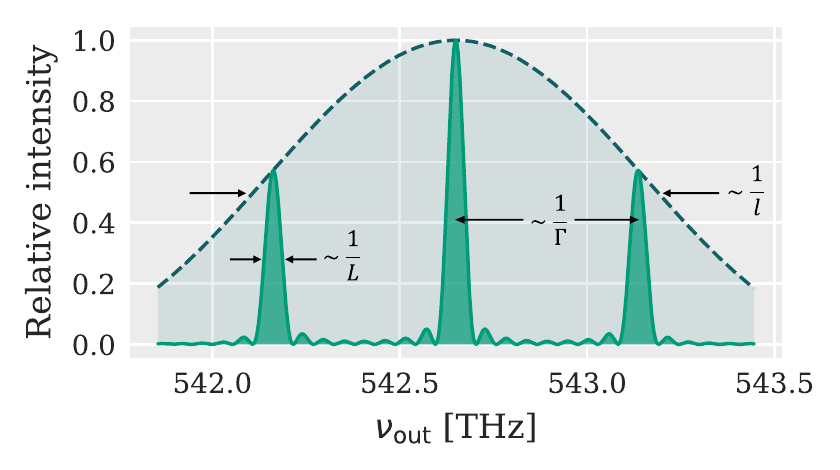}
    \caption{Simulation of the phase-matching intensity showing how the spectral features are related to the super-poling parameters.}
    \label{fig:pm_1D}
    \bigbreak
    \vfill
    \includegraphics{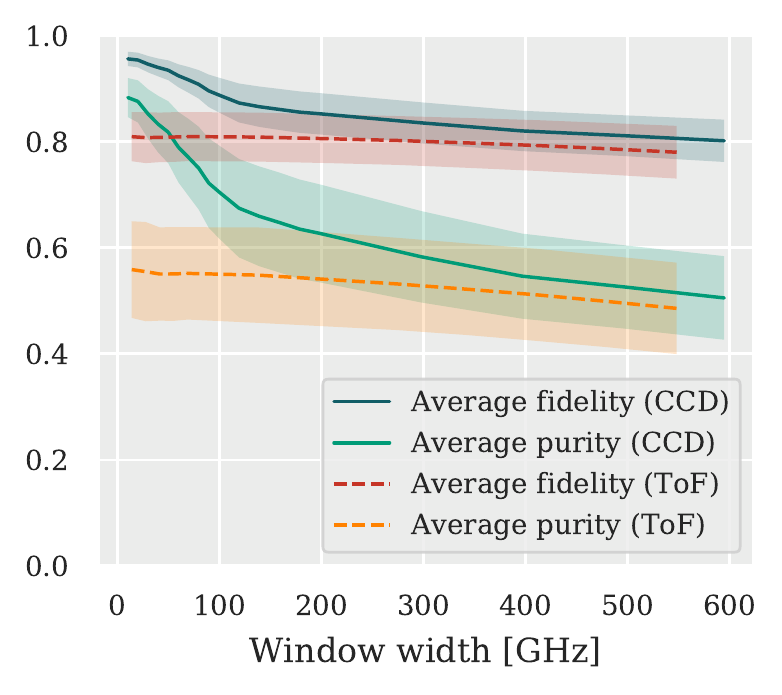}
    \caption{Average purity and fidelity (with corresponding standard deviations) of the POVMs of our implemented five-dimensional decoder for experimental data acquired with the CCD spectrograph (solid line) and the ToF spectrograph (dashed line) as a function of the spectral window around each central frequency.}
    \label{fig:filtering}
\end{figure}

The phase-matching function of the SFG process for a waveguide of length $L$ can be described as a function of the birefringent phase mismatch $\dbu = \beta_\mathrm{out}-\beta_\mathrm{pump}-\beta_\mathrm{in}$, where $\beta_\mathrm{i}$ is the propagation constant of field $i$:
\begin{align}
    \phi_\mathrm{SFG}(\dbu) \propto \int_0^L{g(z) \mathrm{e}^{i\dbu z} \mathrm{d}z} \,,
\end{align}
where $z$ is the position on the main waveguide axis and $g(z)$ is the nonlinearity profile along the waveguide, which is fully determined by the modulation of the poling pattern.
In the range of output wavelengths allowed by energy conservation of the SFG process, the phase-matching function becomes
\begin{align}
    \phi_\mathrm{SFG} \sim \frac{\varGamma}{L} \frac{\mathrm{e}^{i \frac{\dbu L}{2}}}{\mathrm{e}^{i \frac{\dbu \varGamma}{2}}} \frac{\sin(\frac{\dbu L}{2})}{\sin(\frac{\dbu \varGamma}{2})} \, \eta \sinc\left(\frac{\dbp l}{2}\right) \mathrm{e}^{i \frac{\dbp l}{2}} \,,
\end{align}
where we defined $\dbp=\dbu-2\pi/\varLambda$.
This expression describes a train of sinc-shaped peaks modulated by a sinc-shaped envelope. A quick analysis reveals that the physical lengths of the waveguide are related to the spectral parameters: the length of the sample $L$ is inversely proportional to the peak width, the super-poling period $\varGamma$ is inversely proportional to the inter-peak distance, and the length of each poled region $l$ is inversely proportional to the envelope width (Figure \ref{fig:pm_1D}). In particular, the ratio $\varGamma/l = 1/\eta$ defines the number of peaks within the FWHM of the envelope. Therefore, the presented framework allows one to design an mQPG waveguide with the desired number of phase-matching peaks, i.e.\ number of output channels.

In the experimental implementation of the mQPG presented in the main text, we employed a periodically poled Titanium-in-diffused Lithium Niobate waveguide with poling period $\varLambda = \SI{4.32}{\micro\meter}$ operated at a working temperature of \SI{170}{\celsius}. The super-poling parameters were set to $l = \SI{397}{\micro\meter}$ and $\varGamma = \SI{1590}{\micro\meter}$ in order to obtain five output peaks in a wavelength range that would match the bandwidth of the Ti:Sapphire pump $\Delta\nu_\mathrm{Ti:Sa} = \SI{3}{\tera\hertz}$ used in the experiment.

\section{Real mQPG process}
In an ideal mQPG process, each channel is characterized by single-mode operation, that is, provides an output only if the input signal contains the corresponding TM. This ideal behavior is exhibited only if the width of each phase-matching peak is much narrower than that of the corresponding pump pulse \cite{donohue18}. However, as explored in detail in previous work from our group \cite{santandrea19}, a real phase-matching function always displays some degree of imperfection (visible for example in Figure 3 of the main text). Phase-matching imperfections effectively broaden the width of each peak, causing multi-mode behavior in each channel the mQPG process. This multi-modedness then results in output clicks corresponding to TMs that are not present in the input state.

In order to recover single-mode operation in each channel of a real mQPG waveguide, one can exploit the fact that the frequency of the spurious output photons will be slightly offset with respect to the central output frequency of the phase-matching peak. Therefore, one can apply a spectral filter to each output channel in order to discriminate the real counts from the spurious counts arising from the phase-matching imperfections. The method employed in this work is to filter in post-processing the output spectrum acquired by the spectrograph. We define a frequency window around the center of each output peak, and we consider as channel clicks only the counts within this window. The effectiveness of this filtering method is then dependent on the resolution of the spectrograph, that needs to be sufficiently precise to fully distinguish the phase-matching peaks from the surrounding regions.

The definition of resolution for the time-of-flight (ToF) spectrograph requires careful consideration. Since this type of spectrograph is composed of a highly dispersive fiber combined with an avalanche photodiode and a time tagger, one would intuitively define its resolution as that of the time tagger (\SI{1}{\pico\second} in our case) divided by the total chromatic dispersion of the fiber (\SI{217}{\pico\second\per\nano\meter} for our \SI{1}{\kilo\meter}-long fiber). However, one needs to take into account the timing jitter of the ToF setup (\SI{64}{\pico\second} in FWHM) that introduces uncertainty on the arrival time of each photon, hence on its frequency. Therefore, the actual resolution of a ToF spectrograph is given by a combination of these two factors, and in our case is dominated by the latter. This results in a much worse frequency resolution of \SI{300}{\giga\hertz}, which is half as large as the separation between the peaks (\SI{630}{\giga\hertz}).

The effect of this low resolution is visible in Figure \ref{fig:filtering}, which shows the fidelity and the purity of the POVM elements of our decoder as a function of the window width, comparing the measurements obtained with the ToF spectrograph to those obtained with the CCD one (see Table I and Figure 5 from the main text for reference). For the latter, which has a resolution of \SI{30}{\giga\hertz}, one can notice a strong improvement of the measurement quality as the filtering window gets narrower (from right to left), particularly in the average purity of the POVMs. This is due to the progressive elimination of spurious counts in the regions outside the phase-matching peaks. In the case of the ToF spectrograph, on the other hand, the performance improvement is barely noticeable due to its low resolution. Although it is still possible to apply a filtering window narrower than the actual resolution, we see that below \SI{300}{\giga\hertz} the measurement quality is essentially saturated.
The results obtained with the CCD spectrograph are therefore an indication of the high intrinsic quality of our experimental implementation of the mQPG, that could be reached also for the decoder as a whole by solely improving the resolution of the ToF spectrograph.

\bibliography{bibliography}